# Hidden Actions and Hidden Information in Peer Review: A Dynamic Solution

Raphael Mu[*]

July 5, 2024

**Abstract**

We develop a simple model of the scientific peer review process, in which authors of varying ability invest to produce papers of varying quality, and journals evaluate papers based on a noisy signal, choosing to accept or reject each paper. We find that the first-best outcome is the limiting case as the evaluation technology is perfected, even though author type and effort are not known to the journal. Then, we consider the case where journals allow authors to challenge an initial rejection, and find that this approach to peer review yields an outcome closer to the first best relative to the approach that does not allow for such challenges.

---

[*]London School of Economics. email: r.j.mu@lse.ac.uk.

# 1  Introduction

Validation, the process by which scientific writing is evaluated and conferred legitimacy, serves a key function in the research economy. When an author submits their manuscript to a journal for prospective publication, validation is conducted through *peer review*: the journal's editorial staff refer the manuscript to other experts in the author's field, who then advise the editor to accept the manuscript, reject it, or request the author to "revise and resubmit." Stakes are high: authors, who are often employed by universities or research institutions, see their compensation and career progression track closely with their publication record (Hyland, 2012). This has created the loathed "publish-or-perish" paradigm, to which the majority of the scientific community has nonetheless acquiesced. While such a paradigm is lauded for its efficiency (Faria, 2005), it is also argued to have caused distortions in the allocation of attention and funding to research projects and eager exploitation of information asymmetries, which together reduce the efficiency of the system. Since peer review exists at a critical juncture in the research lifecycle, the potential welfare implications of a cheaper and more reliable validation mechanism are large.

We introduce a simple model of the peer review process, in which authors of varying ability produce papers of varying quality, which are then submitted to a journal that decides whether to publish each paper, based on an imperfect signal of the paper's quality. We then augment the model to allow authors to challenge an initial rejection at a cost, and show that this modification can allow the journal to produce a higher-quality corpus. We refer to the scheme in the baseline model by *one-shot review* and the scheme with the option to challenge by *dynamic review*.

It is worth discussing the assumption that authors perfectly observe the quality of their paper, while the journal receives only an imperfect signal. This assumption is made to simplify the analysis, but the results of the model should extend to a more general case where we only assume that authors are *more* informed relative to the journal. We provide some arguments to found this view in the following section.

Notably, our model does not include a fee for the initial submission. In a simpler adverse selection model, where the quality distribution is exogenously given, the journal can ensure an exclusively high-quality corpus by setting a submission fee just high enough so that the individual rationality constraint for authors carrying low-quality papers is not met. That is, where the expected benefit of submitting a low-quality paper is too low to justify incurring the cost of submission. This can be done without negatively affecting authors with high-quality papers, because they expect to receive a positive payoff. However, when the quality distribution is microfounded by the author's research activities, effort choices by authors of all ability levels are sensitive to the submission fee. For clarity, we thus omit the initial submission fee in our model, while still admitting a screening feature through the cost of challenging a rejection. The challenge cost can be interpreted as the effort cost of writing a challenge statement, a response delay imposed by the journal, or a simple fee. This fee, however, would be an externality in the journal's optimization, as it seeks simply to maximize the *impact* of its corpus, as is standard to assume in the literature. Inclusion of an initial submission fee would increase the scope of the model's prediction, to be sure. As we discuss in the following section, submission fees provide a screening function that serves both to encourage self-selection of higher-quality submissions and to reduce referee burden, congestion, and costs in the editorial process.

## 2  Literature Review

There is a small but growing core of literature studying various existing approaches and potential improvements to the peer review process. A main strand considers the use of submission fees and time delays by journals as a device to manage the flow of submissions and improve journal quality (Azar, 2006; Azar, 2015; Barbos, 2013; Besancenot et al., 2011; Besancenot et al., 2014; Cotton, 2013; Muller-Itten, 2021; Tiokhin et al., 2021). Doing so allows journals to cause authors holding lower-quality papers to self-select out of the submission pool, and perhaps even some authors holding higher-quality papers, so as to reduce costs and meet a capacity constraint.

Other approaches to improving the editorial process have been explored. Some researchers recommend refunding the submission fee for accepted papers (Leslie, 2005; Heintzelman and Nocetti, 2009), desk-rejecting less promising papers to reduce burden on referees and editorial staff (Besancenot et al., 2011; Besancenot et al., 2014), optimal tradeoffs between submission fees and time delays, so as not to unfairly advantage wealthy and tenured authors (Cotton, 2013), homogeneous strictness in editorial boards (Besancenot et al., 2012), an opt-in "no-revisions" waiver (Heintzelman and Nocetti, 2009), rules against institutional subsidization of submission fees (Leslie, 2005), limitations on the number of times an author can resubmit a paper (Azar, 2006; Tiokhin et al., 2021), and requiring authors to review in proportion to their submissions (Azar, 2006).

Azar (2015) analyzes a model with similarities to ours, arguing that authors are typically more informed than referees. Journals often use multiple referees out of a desire to obtain a more reliable assessment of the paper. In a study of several international conferences and journals, Dobele (2015) finds that, for about one third of manuscripts, the referees assigned produced conflicting recommendations. Blank (1991) finds that single-blinded and double-blinded review schemes produce consistently different results, and Gans and Shepherd (1994) document highly important articles that were initially rejected by the editorial process. Frey et al. (2009) argue that referees have a conservative bias, and busier academics will tend to delegate drafting of the report to their graduate students, exacerbating the regressive force to orthodox economics. Azar (2015) argues that authors are typically experts in their field, and have spent the most time thinking about their own research, whereas referees may be experts but less knowledgeable, and often spend only a few hours engaging with their assigned manuscripts. Further, as Azar (2015) argues, authors can ask their colleagues for feedback, and it is reasonable to assume that this will typically result in at least as informative a signal as one that would be obtained by a referee. Shah (2022) documents conferences where previous co-authors are often excluded from the peer review process to avoid conflicts of interest. This means that the next-most informed scientists are often passed over by editors when assigning manuscripts for review.

Shah (2022) also documents a conference that allowed for author rebuttal, in which scores changed following about 15-20 percent of rebuttals, with positive changes in twice as many cases as there were negative. Shah (2022) then conveys that rebuttal has a small but significant influence on final scores, especially for marginal papers. These findings contribute a motivation for our model, which we shall now present.

## 3 Model

### 3.1 One-Shot Review

Suppose there is a continuum of authors, having mass 1. Each author has type $\theta_i \in \{\theta_1, \theta_0\}$, with $\theta_1 > \theta_0$, where $\theta_i$ is a measure of productivity, which enters in the author's effort cost. We call authors *skilled* if they have type $\theta_1$, or *unskilled* if they have type $\theta_0$. The proportion of skilled authors in the population is $\mathbb{P}[\theta = \theta_1] = \alpha \in (0, 1)$. Authors begin by conducting research, which produces a paper of quality $q \in \{0, 1\}$, defining $q_i = i$ for notational purposes. We refer to papers of quality 1 as *high-quality* and papers of quality 0 as *low-quality*. The probability that an author produces a high-quality paper, which we call their *success rate*, is a function $y(a)$ of their effort $a$:

$$\mathbb{P}[q = 1|a] = y(a) = 1 - e^{-a}. \tag{1}$$

The author's effort cost is

$$\psi(a; \theta) = \frac{a}{\theta}, \tag{2}$$

which is lower for skilled authors. The author perfectly observes the quality of their paper upon completion. In one-shot review, this assumption is inconsequential, as we do not consider submission fees. However, the author's observation of quality will turn out to affect the author's willingness to challenge a rejection in dynamic review scheme. Next, the author submits the paper to the journal, where an editorial process occurs, yielding a noisy signal of the paper's quality. Here, we treat the journal as an atomic entity, abstracting from the interaction between referees and editors. The journal receives a signal

$$z = q + \varepsilon, \tag{3}$$

where $\varepsilon \sim \mathcal{N}(0, \sigma^2)$. The journal has a chosen evaluation threshold $\bar{z}$, which is set so that papers with signal $z \geq \bar{z}$ are accepted. The probability that a paper with quality $q_i$ is accepted, which we call its *acceptance rate*, is then

$$p_i(\bar{z}) = \mathbb{P}[z \geq \bar{z} | q = q_i] = 1 - \Phi\left(\frac{\bar{z} - q_i}{\sigma}\right), \tag{4}$$

that is, the tail probability for threshold $\bar{z}$. The difference between high-quality and low-quality acceptance rates is

$$p_1(\bar{z}) - p_0(\bar{z}) = \Phi\left(\frac{\bar{z}}{\sigma}\right) - \Phi\left(\frac{\bar{z} - 1}{\sigma}\right). \tag{5}$$

This quantity is important to authors, as it determines the marginal benefit to the author of increasing their success rate, as we shall demonstrate. We normalize the author's benefit of publication to 1. Then, the ex-ante payoff for a type $\theta_i$ author is

$$u_i(a; \bar{z}) = y(a) p_1(\bar{z}) + (1 - y(a)) p_0(\bar{z}) - \psi(a; \theta_i). \tag{6}$$

This equation can be written

$$u_i(a; \bar{z}) = p_0(\bar{z}) + y(a)(p_1(\bar{z}) - p_0(\bar{z})) - \psi(a; \theta_i),$$

revealing that the marginal benefit of effort is the marginal success rate scaled by the difference in acceptance rates. An author of type $\theta_i$ hence chooses an effort

$$a_i(\bar{z}) = \operatorname*{argmax}_a u_i(a; \bar{z}), \qquad (7)$$

which implies a success rate $y_i(\bar{z}) = y(a_i(\bar{z}))$. The total share of high-quality papers produced by authors of all types will be

$$\beta(\bar{z}) = \alpha y_1(\bar{z}) + (1-\alpha) y_0(\bar{z}), \qquad (8)$$

and the quantity of high-quality papers published in the journal will be

$$X(\bar{z}) = \beta(\bar{z}) p_1(\bar{z}), \qquad (9)$$

which we call the *impact function*. This is the quantity that the journal seeks to maximize. The total quantity of papers published will be

$$Y(\bar{z}) = \beta(\bar{z}) p_1(\bar{z}) + (1 - \beta(\bar{z})) p_0(\bar{z}), \qquad (10)$$

which we call the *yield function*.

**Proposition 1** *In one-shot review, the quantity of high-quality papers produced can be written as*

$$\beta(\bar{z}) = 1 - \frac{\bar{c}}{p_1(\bar{z}) - p_0(\bar{z})}, \qquad (11)$$

*and the yield function can be written as*

$$Y(\bar{z}) = p_1(\bar{z}) - \bar{c}, \qquad (12)$$

*where*

$$\bar{c} = \frac{\alpha}{\theta_1} + \frac{1-\alpha}{\theta_0}. \tag{13}$$

We call $\bar{c}$ the *average cost* of research, because the marginal cost of effort for an agent of type $\theta_i$ is

$$\psi_a(a; \theta_i) = \frac{1}{\theta_i}.$$

Observe that, since $p_1(\bar{z})$ strictly decreases in $\bar{z}$, so does $Y(\bar{z})$, the journal's yield. For a given batch of completed papers, the journal can maximize its impact by simply publishing every paper, which is the limiting case as $\bar{z} \to -\infty$. However, authors' incentive compatibility constraints would dictate that they exert zero effort, as even low-quality papers would always be accepted. Thus, we expect the journal to choose some degree of selectivity, in order to incentivize production of high-quality papers. Additionally, we assume that the journal has a capacity constraint $n \in (0,1)$, meaning that the journal can publish at most $n$ papers. Thus, the journal chooses an evaluation threshold

$$\bar{z} = \underset{\bar{z}}{\operatorname{argmax}} \, X(\bar{z}), \tag{14}$$

subject to

$$Y(\bar{z}) \leq n. \tag{15}$$

## 3.2 Dynamic Review

In one-shot review, the publication outcome is bound to the journal's initial judgment. Suppose now that authors may challenge an initial rejection at cost $\kappa$. This will cause the journal to reevaluate the paper, acquiring a new signal. We considered several ways to model this, deciding that the journal should simply redraw an evaluation from the signal distribution. This can be interpreted as simply choosing a second referee (or team of referees) to evaluate the paper. While it may be less efficient for an editor to ignore the first signal during reevaluation, it simplifies the model by making the initial and challenge acceptance rates independent and thus identical, for a given paper quality. That is, both acceptance rates are $p_i(\bar{z})$. We argue that this decision does not affect the qualitative results of the model, and proceed. The probability that a paper of quality $q_i$ is accepted, either initially or after an optional challenge, which we call the *total acceptance rate*, is

$$\hat{p}_i(\bar{z}, s_i) = p_i(\bar{z}) + s_i(1 - p_i(\bar{z}))p_i(\bar{z}), \tag{16}$$

where $s_i$ is the author's choice of whether to challenge a rejection of that paper. The author's *virtual acceptance rate*, which takes into account the cost of challenging a rejection, is

$$\tilde{p}_i(\bar{z}, \kappa, s_i) = \hat{p}_i(\bar{z}, s_i) - s_i(1 - p_i(\bar{z}))\kappa. \tag{17}$$

Thus, the author's ex-ante payoff is

$$\hat{u}_i(x; \nu) = y(a)\tilde{p}_1(\nu, s_1) + (1 - y(a))\tilde{p}_0(\nu, s_0) - \psi(a; \theta_i), \tag{18}$$

where $x = (a, s_1, s_0)$ and $\nu = (\bar{z}, \kappa)$. The author then chooses

$$x_i(\nu) = \underset{x}{\operatorname{argmax}}\ \hat{u}_i(x; \nu), \tag{19}$$

which implies a total success rate $\hat{y}_i(\nu) = y(a_i(\nu))$. As we show in the following section, authors will symmetrically choose $s_{ji}(\nu) = s_j(\nu)$, where $i$ indicates author type and $j$ indicates paper quality. Similarly to the one-shot case, the quantity of high-quality papers is

$$\hat{\beta}(\nu) = \alpha\hat{y}_1(\nu) + (1 - \alpha)\hat{y}_0(\nu), \tag{20}$$

the journal's impact will be

$$\hat{X}(\nu) = \hat{\beta}(\nu)\hat{p}_1(\bar{z}, s_1(\nu)), \tag{21}$$

and the journal's yield will be

$$\hat{Y}(\nu) = \hat{\beta}(\nu)\hat{p}_1(\bar{z}, s_1(\nu)) + (1 - \hat{\beta}(\nu))\hat{p}_0(\bar{z}, s_0(\nu)). \qquad (22)$$

**Proposition 2** *In dynamic review, the quantity of high-quality papers produced can be written as*

$$\hat{\beta}(\nu) = 1 - \frac{\bar{c}}{\tilde{p}_1(\nu) - \tilde{p}_0(\nu)}, \qquad (23)$$

*and the yield function can be written as*

$$\hat{Y}(\nu) = Y(\bar{z}) + s_1(\nu)(1 - p_1(\bar{z}))p_1(\bar{z}) - \bar{c}\frac{\tau(\nu)}{\tilde{p}_1(\bar{z}, s_1(\nu)) - \tilde{p}_0(\bar{z}, s_0(\nu))}, \qquad (24)$$

*where*

$$\tau(\nu) = [s_1(\nu)(1 - p_1(\bar{z})) - s_0(\nu)(1 - p_0(\bar{z}))]\kappa. \qquad (25)$$

In dynamic review, the journal chooses a policy

$$\nu = \underset{\nu}{\operatorname{argmax}}\ X(\nu), \qquad (26)$$

subject to

$$\hat{Y}(\nu) \leq n. \qquad (27)$$

We now proceed to solve each case of our model.

## 4 Analysis

### 4.1 First Best

To characterize the first-best outcome, we consider an environment in which the journal is able to verify author type and effort. This can be implemented through a contract where the journal commits to accept the author's manuscript if and only if the paper is of high quality and the author chooses the first-best effort $a_i^*$. In the first best, the journal dictates

$$(a_1, a_0) = \underset{a_1, a_0}{\mathrm{argmax}} \; \{\beta(a_1, a_0) - (\alpha \psi(a_1; \theta_1) + (1-\alpha)\psi(a_0; \theta_0))\}, \tag{28}$$

subject to $\beta(a_1, a_0) \leq n$.

**Theorem 1** *In the first best, if $n \geq 1 - \bar{c}$, an author of type $\theta_i$ chooses effort*

$$a_i^* = \ln \theta_i, \tag{29}$$

*with success rate*

$$y_i = 1 - \frac{1}{\theta_i}, \tag{30}$$

*and the journal has impact*

$$\beta(a_1, a_0) = 1 - \bar{c}. \tag{31}$$

*If $n < 1 - \bar{c}$, an author of type $\theta_i$ chooses effort*

$$a_i^* = \ln \theta_i + \ln \frac{\bar{c}}{1-n}, \tag{32}$$

*with success rate*

$$y_i = 1 - \frac{1}{\theta_i} \frac{1-n}{\bar{c}}, \tag{33}$$

*and the journal has impact*

$$\beta(a_1, a_0) = n. \tag{34}$$

We see that if the journal's capacity constraint binds, then the journal sets lower effort levels for both authors (as $\bar{c} < 1-n$), saving on costs from producing the $1-\bar{c}-n$ high-quality papers that go to waste. Interestingly, so long as quality alone is verifiable, the journal is still able to achieve the first best.

**Theorem 2** *If the journal can verify paper quality, the principal can implement the first*

*best by always accepting high-quality papers if $n \geq 1 - \bar{c}$, or by accepting high-quality papers with probability $\frac{\bar{c}}{1-n}$ if $n < 1 - \bar{c}$.*

This suggests that, as referee precision is perfected, the first-best can be achieved, even where there is an information asymmetry arising from the journal's lack of knowledge of the author's type and effort.

## 4.2 Equilibrium under One-Shot Review

In one-shot review, authors of type $\theta_i$ choose effort

$$a_i(\bar{z}) = \underset{a}{\operatorname{argmax}}\ u_i(a; \bar{z}), \tag{7}$$

with first-order condition

$$y_a(a_i(\bar{z})) = \frac{1}{\theta_i(p_1(\bar{z}) - p_0(\bar{z}))}, \tag{35}$$

that is,

$$a_i(\bar{z}) = \ln(\theta_i(p_1(\bar{z}) - p_0(\bar{z}))). \tag{36}$$

Since $p_1(\bar{z}), p_0(\bar{z}) \in (0,1)$ for finite values of $\bar{z}$, we see that equilibrium effort is strictly lower than the first-best effort. The validity condition $y_i(\bar{z}) \geq 0$ necessitates

$$y_a(a) \geq \frac{1}{\theta_i(p_1(\bar{z}) - p_0(\bar{z}))} \tag{37}$$

for some $a \geq 0$, which is satisfied for

$$y_a(0) \geq \frac{1}{\theta_0(p_1(\bar{z}) - p_0(\bar{z}))}, \tag{38}$$

since $\theta_0 < \theta_1$, and which is maximally weak since $y_{aa} < 0$.

The journal solves

$$\bar{z} = \underset{\bar{z}}{\operatorname{argmax}}\ X(\bar{z}), \tag{14}$$

subject to
$$Y(\bar{z}) \leq n. \tag{15}$$

The first-order condition for the unconstrained optimization is

$$(V - U)^2 u = \bar{c}((1 - U)v - (1 - V)u),$$

where

$$U = \Phi\left(\frac{\bar{z} - 1}{\sigma}\right), \quad V = \Phi\left(\frac{\bar{z}}{\sigma}\right),$$
$$u = \phi\left(\frac{\bar{z} - 1}{\sigma}\right), \quad v = \phi\left(\frac{\bar{z}}{\sigma}\right),$$

which has no analytical solution. Borrowing from Azar (2015), we instead use the approach of setting $n = Y$ in equilibrium. In Azar (2015), the quality distribution is exogenously given, so the journal maximizes impact by setting the evaluation threshold as low as possible without having to turn away any papers that exceed the evaluation threshold due to capacity constraints. In our case, impact is non-monotone quasiconcave. However, it has a unique optimizer $\bar{z}^*$, and $X(\bar{z})$ is strictly decreasing in $\bar{z}$ for $\bar{z} > \bar{z}^*$. Since $Y(\bar{z})$ also strictly decreases in $\bar{z}$, as long as the unconstrained optimum is unattainable due to capacity constraints, our journal would like to maximize yield, subject to not turning away any papers with successful evaluations. In other words, the journal's optimization problem binds at $n = Y(\bar{z})$, so long as $n < Y(\bar{z}^*)$. This is a reasonable assumption for journals in large disciplines. Since the supply of papers typically exceeds journals' capacity to publish, if there are $k$ journals in a discipline, then we would expect $n < \frac{1}{k}$ for a typical journal in a competitive field. This is evidenced by the fact that many journals experience a high degree of demand and are quite overburdened in practice. We thus restrict our attention to this case. If the journal sets $\bar{z}$ so that $n = Y(\bar{z})$, we have

$$\bar{z} = \sigma \Phi^{-1}(1 - \bar{c} - n) + 1 \tag{39}$$

in equilibrium, by Equation 12. Since $\bar{c} + n = p_1(\bar{z}) \in (0, 1)$, the quantile is well-defined. The intuition is as follows.

As referees become more precise ($\sigma$ decreases), high-quality papers are less likely to meet an evaluation threshold that is above their quality ($\bar{z} > 1$). However, low-quality papers are even less likely to do so, as a consequence of the monotone likelihood ratio property. Thus, the evaluation threshold required to establish a sufficient signal of high quality lowers, because better precision means that less journal space is spent on low-quality papers. If $\bar{z} < 1$, high-quality papers will be more likely to meet an evaluation threshold following an improvement in precision, while low-quality papers will be less likely. Thus, the journal can afford to set a higher evaluation threshold, improving the likelihood ratio. As the paper's capacity grows ($n$ increases), the evaluation threshold decreases. This is because the journal can afford to be less stringent and still expect to receive the highest-quality papers. As skilled or unskilled authors become more productive, or as the proportion of skilled authors $\alpha$ increases, average cost $\bar{c}$ decreases. The evaluation threshold increases in turn, because the journal expects a more competent population that is able to meet higher standards.

## 4.3 Equilibrium under Dynamic Review

In dynamic review, authors of type $\theta_i$ choose effort

$$x_i(\nu) = \operatorname*{argmax}_{x} \hat{u}_i(x; \nu), \tag{19}$$

with first-order conditions

$$a_i(\nu) = \ln(\theta_i [\tilde{p}_1(\nu, s_{1i}(\nu)) - \tilde{p}_0(\nu, s_{0i}(\nu))]) \tag{40}$$

$$s_{ji}(\nu) = \mathbb{I}[p_j(\bar{z}) \geq \kappa], \tag{41}$$

assuming that, in the marginal case where an author is indifferent as to whether to challenge, the author will attempt the challenge. Since the optimal choice of whether to challenge a rejection does not depend on author type, authors symmetrically set $s_{ji}(\nu) = s_j(\nu)$ in equilibrium. The choice of $s_j(\nu)$ is nonincreasing in both $\bar{z}$ and $\kappa$, as raising the evaluation threshold or raising the cost of challenge both tighten the constraint $p_j(\bar{z}) \geq \kappa$. Then, since $p_1(\bar{z}) > p_0(\bar{z})$, we classify three kinds of $(\bar{z}, \kappa)$-policy, based on how authors best-respond with $s_j$.

If $p_1(\bar{z}) < \kappa$, then $s_1 = s_0 = 0$, and we say that $(\bar{z}, \kappa)$ is *hawkish*. Authors never challenge a rejection, no matter the quality of their paper. For such policies, the challenge feature is never used and thus dynamic review produces the same outcome as in one-shot review.

If $p_0(\bar{z}) < \kappa \leq p_1(\bar{z})$, then $s_1 = 1$ and $s_0 = 0$, and we say that $(\bar{z}, \kappa)$ is *moderate*. For these policies, effort increases above the one-shot effort as $\kappa$ decreases. This is because a cheaper challenge means that authors can extract a higher expected benefit from their effort. As $\bar{z}$ decreases, effort also increases above the one-shot effort because a challenge is more likely to succeed, although sometimes only up to a point. This is because too low of an evaluation threshold will reduce the difference in acceptance rates between high-quality and low-quality papers, reducing the incentive to work.

If $\kappa \leq p_0(\bar{z})$, then $s_1 = s_0 = 1$, and we say that $(\bar{z}, \kappa)$ is *dovish*. Among dovish policies, effort decreases in dovishness. This is because, from the evaluation perspective, a challenge is too much more likely to lead to an acceptance, while from a cost perspective, it is too much cheaper to make a challenge. Either way, authors rely on challenging rather than working harder. Under extremely dovish policies, effort actually falls below that of the one-shot case.

Finally, the validity condition $y_i(\nu) \geq 0$ necessitates

$$y_a(a) \geq \frac{1}{\theta_i(\tilde{p}_1(\nu, s_1(\nu)) - \tilde{p}_0(\nu, s_0(\nu)))} \tag{42}$$

for some $a \geq 0$, which is satisfied for

$$y_a(0) \geq \frac{1}{\theta_0(\tilde{p}_1(\nu, s_1(\nu)) - \tilde{p}_0(\nu, s_0(\nu)))}, \tag{43}$$

since $\theta_0 < \theta_1$, and which is maximally weak since $y_{aa} < 0$. We now discuss some statics on author behavior.

As $\sigma^2$ increases, the gap between yield and impact grows larger. Efforts and success rates decrease under intermediate policies, while they increase under extreme policies.

For $\sigma^2 \to 0$, authors with high-quality papers are always willing to challenge for $\bar{z} < 1$, and authors with low-quality papers are always willing to challenge for $\bar{z} < 0$. Both of these constraints weaken as $\sigma^2$ increases from 0, as for any given threshold above the paper's true quality, there is a higher chance of "leaping over" the threshold. In particular, the level below which an evaluation threshold is considered dovish increases.

As $\alpha$ or either $\theta_i$ increases, the high-quality share $\beta$ will increase, and thus both yield and impact increase.

As $\kappa$ increases, efforts and success rates decrease under moderate policies, while they increase under dovish policies. The former is because the virtual acceptance rate for high-quality papers goes down, reducing the incentive to exert effort. The latter is because the probability of a high-quality rejection is lower than that of a low-quality rejection, and so if the author always challenges, a higher challenge cost results in a higher difference in virtual acceptance rates and thus a greater incentive to exert effort.

Next, we turn to the journal's choice of $\nu = (\bar{z}, \kappa)$. The journal solves

$$\nu = \underset{\nu}{\operatorname{argmax}}\, X(\nu), \tag{26}$$

subject to

$$Y(\nu) \leq n. \tag{27}$$

Once again, we argue that journals are small enough for the capacity constraint to bind. The following result allows us to characterize the equilibrium.

**Theorem 3** *For a given evaluation threshold, the yield under dynamic review is at least as great as in one-shot review. That is, for all $\nu = (\bar{z}, \kappa)$,*

$$\hat{Y}(\nu) \geq Y(\bar{z}). \tag{44}$$

*Further, for non-hawkish policies, the inequality is strict.*

Let $\Delta = \hat{Y}(\nu) - Y(\bar{z}) \geq 0$. Then, by Equation 12 and choosing $n = \hat{Y}(\nu)$, we have

$$\bar{z} = \sigma \Phi^{-1}(1 - \bar{c} - n + \Delta) + 1. \tag{45}$$

For a given level of $\kappa$, it is possible for the most capacity-constrained journals to set a hawkish policy. However, since the journal can always incentivize greater effort by reducing $\kappa$, the journal will choose some non-hawkish policy, with $\kappa < 1$. This means that $\Delta > 0$, and so the journal will set a higher evaluation threshold than in the one-shot case. It is interesting that the journal should do so, as a lower probability of acceptance can result in lower efforts and success rates. However, giving authors the prospect of challenging a rejection results in a large enough increase in efforts and success rates to offset the reduction due to stricter evaluations. For moderate policies, this guarantees a higher ex-ante effort from authors as well, bringing the journal closer to the first-best outcome.

## 5 Extensions

In this section, we discuss a number of potentially fruitful ways of extending the model, to accommodate additional structure, dynamics, or heterogeneity. For example, we speculate that by adding the option to challenge rejections multiple times, the journal will be able to obtain more information about paper quality and achieve a higher impact than under single-challenge dynamic review. Also, initial submission fees and time delays will give the journal an additional screening device, and importantly, allow it to reduce referee and editor costs. From there, we may gain a fuller view of the impact of heterogeneity in wealth or patience among authors. By expanding the set of possible author abilities and paper qualities, perhaps to a continuous distribution, it will be possible to find richer dynamics than with our bivariate Bernoulli case. Also, we assumed that the author's publication benefit is given exogenously. One may endogenize this value, so that author benefit increases in the impact, average quality, or notoriety of a journal. Further, one can introduce competition between journals, based on this benefit as well as expected time and pecuniary costs. Another dimension of heterogeneity can arise from the specialization of authors and journals in different fields. It would be interesting to study a case where journals compete in a Hotelling spatial model. Journals may choose some optimal path over which to horizontally differentiate over time, reacting to shocks such as research subsidies, breakthroughs, and journal closures due to bankruptcy. Referees may also have reputational concerns: an editor may differentially assign opportunities to referees, or perhaps even offer them a seat on the editorial board, depending on their track record of precision. Finally, it may be proper to close the distance between referee and author utility, since after all, it is the authors themselves who review those papers seeking to be published.

# 6   Conclusion

We have introduced a simple model of the peer review process and conducted a rudimentary analysis, yielding several straightforward results, as well as a ream of ancillary predictions. First, so long as a journal can perfectly verify the quality of a paper, it is able to achieve the first-best outcome, even if it is unaware of authors' abilities or efforts. This is the limiting case as referee precision is perfected, suggesting that investment into the evaluation technology, perhaps directly by improving referee precision through better training and matching, or indirectly through organizational strategies, should be a long-term priority. Second, we learned that dynamic review enables the journal to improve its impact and draw nearer to the first-best outcome. This is because the opportunity to challenge rejection gives the journal an additional observation of paper quality, and also preferentially improves the success rates of skilled authors. We have also uncovered a number of useful statics, such as the direct mechanical impact of author costs on yield, the desire of the journal to set an intermediate evaluation threshold, the local benefit of increasing journal capacity, and the influence of the evaluation threshold, challenge cost, and referee precision on a host of decisions. As discussed in the section above, there is ample opportunity to extend the model, so as to give richer predictions of peer review. As time progresses, we expect more of these questions to be answered, and for the scientific process to gradually improve.

# Appendix. Proofs of Theorems and Propositions

## Proof of Proposition 1

Suppressing arguments for clarity, begin with

$$Y = \beta p_1 + (1-\beta) p_0$$

$$\beta = \alpha y_1 + (1-\alpha) y_0$$

$$y_i = 1 - \frac{1}{\theta_i(p_1 - p_0)}.$$

Then,

$$\beta = \alpha y_1 + (1-\alpha) y_0$$

$$= \alpha \left(1 - \frac{1}{\theta_1(p_1 - p_0)}\right) + (1-\alpha)\left(1 - \frac{1}{\theta_0(p_1 - p_0)}\right)$$

$$= 1 - \frac{\bar{c}}{p_1 - p_0},$$

where

$$\bar{c} = \frac{\alpha}{\theta_1} + \frac{1-\alpha}{\theta_0}.$$

Combining with the expression for $Y$ gives

$$Y = p_0 + \beta(p_1 - p_0)$$

$$= p_1 - \bar{c},$$

as desired.

## Proof of Proposition 2

We take an approach similar to the proof for Proposition 1, once again suppressing arguments for clarity. Begin with

$$\hat{Y} = \hat{\beta}\hat{p}_1 + (1-\hat{\beta})\hat{p}_0$$

$$\hat{\beta} = \alpha\hat{y}_1 + (1-\alpha)\hat{y}_0$$

$$\hat{y}_i = 1 - \frac{1}{\theta_i(\tilde{p}_1 - \tilde{p}_0)}.$$

Then, observe that

$$\hat{\beta} = \alpha\left(1 - \frac{1}{\theta_1(\tilde{p}_1 - \tilde{p}_0)}\right) + (1-\alpha)\left(1 - \frac{1}{\theta_0(\tilde{p}_1 - \tilde{p}_0)}\right)$$

$$= 1 - \frac{\bar{c}}{\tilde{p}_1 - \tilde{p}_0}.$$

Next, define

$$\tau = [s_1(1-p_1) - s_0(1-p_0)]\kappa,$$

so that

$$\tilde{p}_1 - \tilde{p}_0 = \hat{p}_1 - \hat{p}_0 - \tau.$$

Then,

$$\hat{\beta}(\hat{p}_1 - \hat{p}_0) = \hat{\beta}(\tilde{p}_1 - \tilde{p}_0 + \tau)$$
$$= \tilde{p}_1 - \tilde{p}_0 - \bar{c} + \hat{\beta}\tau$$
$$= \hat{p}_1 - \hat{p}_0 - \bar{c} - (1 - \hat{\beta})\tau$$
$$= \hat{p}_1 - \hat{p}_0 - \bar{c} - \frac{\bar{c}}{\tilde{p}_1 - \tilde{p}_0}\tau$$
$$= \hat{p}_1 - \hat{p}_0 - \bar{c}\left(1 + \frac{\tau}{\tilde{p}_1 - \tilde{p}_0}\right)$$
$$= \hat{p}_1 - \hat{p}_0 - \bar{c}\left(\frac{\hat{p}_1 - \hat{p}_0}{\tilde{p}_1 - \tilde{p}_0}\right).$$

Combining with the expression for $\hat{Y}$ gives

$$\hat{Y} = \hat{p}_0 + \hat{\beta}(\hat{p}_1 - \hat{p}_0)$$
$$= \hat{p}_1 - \bar{c}\left(\frac{\hat{p}_1 - \hat{p}_0}{\tilde{p}_1 - \tilde{p}_0}\right)$$
$$= Y + s_1(1 - p_1)p_1 - \bar{c}\left(\frac{\tau}{\tilde{p}_1 - \tilde{p}_0}\right),$$

as desired.

## Proof of Theorem 1

The journal solves

$$(a_1, a_0) = \underset{a_1, a_0}{\operatorname{argmax}}\ \{\beta(a_1, a_0) - (\alpha\psi(a_1; \theta_1) + (1-\alpha)\psi(a_0; \theta_0))\},$$

subject to

$$\beta(a_1, a_0) \leq n.$$

The Lagrangian function for this optimization is

$$\mathcal{L} = \beta(a_1, a_0) - (\alpha\psi(a_1; \theta_1) + (1-\alpha)\psi_0(a_0; \theta_0)) - \lambda[\beta(a_1, a_0) - n]$$
$$= \alpha[(1-\lambda)y_1(a_1) - \psi(a_1; \theta_1)] + (1-\alpha)[(1-\lambda)y_0(a_0) - \psi(a_0; \theta_0)] + \lambda n,$$

with first-order conditions

$$\frac{\partial \mathcal{L}}{\partial a_1} = \alpha\left((1-\lambda)e^{-a_1} - \frac{1}{\theta_1}\right) = 0$$
$$\frac{\partial \mathcal{L}}{\partial a_0} = (1-\alpha)\left((1-\lambda)e^{-a_0} - \frac{1}{\theta_0}\right) = 0$$

and

$$\frac{\partial \mathcal{L}}{\partial \lambda} = n - \beta.$$

These imply

$$e^{a_i} = \theta_i(1-\lambda)$$

and

$$\frac{e^{a_1}}{e^{a_0}} = \frac{\theta_1}{\theta_0}.$$

The unconstrained maximum, where $\lambda = 0$, is

$$a_i = \ln \theta_i,$$

so that

$$\beta = \alpha(1 - e^{-a_1}) + (1 - \alpha)(1 - e^{-a_0})$$

$$= 1 - (\alpha e^{-a_1} + (1 - \alpha)e^{-a_0})$$

$$= 1 - \bar{c}.$$

An author of type $\theta_i$ thus has success rate

$$y_i = 1 - \frac{1}{\theta_i}.$$

If $n < 1 - \bar{c}$, the capacity constraint binds. Then, since

$$e^{-a_0} = e^{-a_1} \frac{\theta_1}{\theta_0},$$

we have

$$n = \beta$$
$$= \alpha(1 - e^{-a_1}) + (1-\alpha)(1 - e^{-a_0})$$
$$= \alpha(1 - e^{-a_1}) + (1-\alpha)\left(1 - e^{-a_1}\frac{\theta_1}{\theta_0}\right)$$
$$= 1 - \left(\alpha + (1-\alpha)\frac{\theta_1}{\theta_0}\right)e^{-a_1}$$
$$= 1 - \bar{c}\theta_1 e^{-a_1}$$
$$e^{a_1} = \frac{\bar{c}\theta_1}{1-n}.$$

This means that an author of type $\theta_i$ chooses effort

$$a_i^* = \ln \theta_i + \ln \frac{\bar{c}}{1-n},$$

with success rate

$$y_i = 1 - \frac{1}{\theta_i}\frac{1-n}{\bar{c}},$$

where both are lower than in the unconstrained case, since $\bar{c} < 1-n$. The Lagrange multiplier in the constrained case is

$$\lambda = \frac{1 - \bar{c} - n}{1-n},$$

which is the ratio of wasted high-quality papers to the journal's unfulfillable demand. This can be interpreted as the marginal improvement in impact that the journal can capture by increasing its capacity.

## Proof of Theorem 2

Suppose the journal can verify paper quality and $n \geq 1 - \bar{c}$. Then, if the journal commits to only accepting high-quality papers, an author of type $\theta_i$ chooses

$$a_i = \underset{a}{\operatorname{argmax}} \ \{y(a) - \psi(a; \theta_i)\},$$

with first-order condition

$$a_i = \ln \theta_i,$$

implying success rate

$$y_i = 1 - \frac{1}{\theta_i}.$$

The journal's impact will be

$$\begin{aligned} \beta &= \alpha y_1 + (1 - \alpha) y_0 \\ &= \alpha \left(1 - \frac{1}{\theta_1}\right) + (1 - \alpha) \left(1 - \frac{1}{\theta_0}\right) \\ &= 1 - \bar{c}, \end{aligned}$$

which is the first-best outcome. Now suppose that $n < 1 - \bar{c}$. Then, if the journal commits to accepting high-quality papers with probability $\frac{\bar{c}}{1-n}$, an author of type $\theta_i$ chooses

$$a_i = \underset{a}{\operatorname{argmax}} \ \left\{\frac{\bar{c}}{1-n} y(a) - \psi(a; \theta_i)\right\},$$

with first-order condition

$$a_i = \ln \theta_i + \ln \frac{\bar{c}}{1-n},$$

implying success rate

$$y_i = 1 - \frac{1}{\theta_i}\frac{1-n}{\bar{c}}.$$

The journal's impact will be

$$\begin{aligned}\beta &= \alpha\left(1 - \frac{1}{\theta_1}\frac{1-n}{\bar{c}}\right) + (1-\alpha)\left(1 - \frac{1}{\theta_0}\frac{1-n}{\bar{c}}\right)\\ &= 1 - \frac{1-n}{\bar{c}}\bar{c}\\ &= n,\end{aligned}$$

which is, once again, the first-best outcome.

## Proof of Theorem 3

We suppress arguments for clarity. Begin with

$$\hat{Y} - Y = s_1(1-p_1)p_1 - \bar{c}\kappa\frac{s_1(1-p_1) - s_0(1-p_0)}{\tilde{p}_1 - \tilde{p}_0}$$

from Proposition 2. For $s_1 = s_0 = 0$, we have $\hat{Y} = Y$. Otherwise, from the validity assumption Equation 43, we have

$$\tilde{p}_1 - \tilde{p}_0 \geq \frac{1}{\theta_0} > \bar{c}.$$

Since $s_1 = 1$, we know that $p_1 \geq \kappa$, and so $p_1(\tilde{p}_1 - \tilde{p}_0) > \bar{c}\kappa$. Multiplying both sides by

$(1-p_1)p_1$ and dividing by $\tilde{p}_1 - \tilde{p}_0$, we have

$$\begin{aligned}(1-p_1)p_1 &> \bar{c}\frac{(1-p_1)p_1}{\tilde{p}_1 - \tilde{p}_0} \\ &> \bar{c}\kappa\frac{1-p_1}{\tilde{p}_1 - \tilde{p}_0} \\ &> \bar{c}\kappa\frac{1-p_1 - s_0(1-p_0)}{\tilde{p}_1 - \tilde{p}_0},\end{aligned}$$

from which we conclude $\hat{Y} > Y$.